# Causality in classical electrodynamics

**C. M. Savage**
**Department of Quantum Science, The Australian National University**
The Physics Teacher, in press.
craig.savage@anu.edu.au

Causality in electrodynamics is a subject of some confusion, especially regarding the application of Faraday's law and the Ampere-Maxwell law.[1,2,3] This has led to the suggestion that we should not teach students that electric and magnetic fields can cause each other,[1] but rather focus on charges and currents as the causal agents.[3] In this paper I argue that fields have equal status as casual agents, and that we should teach this. Following a discussion of causality in classical physics I will use a numerical solution of Maxwell's equations to inform a field based causal explanation in electrodynamics.

In classical physics the theoretical entities, particles and fields, may be regarded as actually existing in the world – a view called "realism". Hence classical physical explanations may be given in terms of the continuous propagation through space of causes and effects. Such an explanation scheme is called "local causality". In quantum physics the situation is different, as it requires neither realism nor locality. This paper is confined to explanations in classical physics.

Causality is the key physics concept for explaining change. In mechanics changes in momentum $m\vec{v}$ are caused by forces $\vec{F}$, according to Newton's second law:

$$\frac{dm\vec{v}}{dt} = \vec{F}. \tag{1}$$

The derivative describes what is changing in time *t* and the right hand side describes the cause of that change. For Newton, gravitational forces could act remotely. However the introduction of the concept of the field into physics replaced such action at a distance with local causality.

In electromagnetism Faraday's law and the Ampere-Maxwell law describe changes in the magnetic $\vec{B}$ and electric $\vec{E}$ fields:

$$\frac{\partial \vec{B}}{\partial t} = -\nabla \times \vec{E}, \qquad \text{Faraday's law ;} \tag{2}$$

$$\frac{\partial \vec{E}}{\partial t} = c^2 \nabla \times \vec{B} - \varepsilon_0^{-1} \vec{J}, \quad \text{Ampere-Maxwell law.} \tag{3}$$

*c* is the speed of light, $\varepsilon_0$ is the vacuum permittivity, and $\vec{J}$ is the current density. Following Bohr,[4] it is reasonable to interpret these equations by analogy with equation (1): the derivatives describe what is changing and the causes of the changes are on the right hand side. Faraday's law then says that the local cause of change in the magnetic field is spatial variation of the electric field, specifically its curl. This inverts the interpretation, refuted by Hill,[1] that the changing magnetic field causes an electric field. The Ampere-Maxwell law says that the local causes of change in the electric field are the curl of the magnetic field and the current density. Together, these equations imply that the electric and magnetic fields cause each other. They might be called the "causal equations", as the other two of Maxwell's equations do not involve time derivatives, and may be derived from equations (2) and (3) by taking divergences, integrating over time, and using charge conservation.

In contrast, Hill [1,2] and Jefimenko [3] claim that causes must precede effects and that therefore equations (2) and (3) are causally ambiguous because the purported effect (left hand side) is simultaneous with the purported cause (right hand side). However if this were the case then we would be left without a local causal explanation for how an electromagnetic wave enters a region $A$ where the fields are initially zero. The local cause of the field growth at a point in $A$ is the spatial variation of the fields at adjacent points due to the incoming wave.

A closely related example is the propagation of a sinusoidally varying, propagating plane electromagnetic wave, in which the electric and magnetic fields are in phase. At the nodes, where the fields are zero, the curls of the fields are greatest, which causes the field changes adjacent to the nodes.

Causality extends the concept of correlation by adding an explanation for why an effect is a consequence of its cause, rather than simply always occurring together with it. For example, Faraday observed a correlation between switching currents on and off and the occurrence of induced currents in nearby circuits. Maxwell's theory of electromagnetism provided the local causal explanation as an electromagnetic field propagating from the switched circuit to the induced current's circuit. But when there is a causal chain one may also consider "retarded causes" that occur earlier in time than the local causes. For example, although the local cause of the induced current is an electric field, the retarded cause is the switching of the current.

In classical mechanics causality is uncontroversial. However as Eric Hill has recently pointed out in this journal the situation in electromagnetism is confused.[1,2] After showing that Faraday's law does not support the claim that a time-varying magnetic field causes an electric field he concludes that we should not teach students that "fields have the unphysical capacity to source each other".[1] Rather, he suggests focusing on the retarded causes of the fields from charges and currents, as expressed by the Jefimenko equations.[3]

However by focusing on retarded causes we risk undervaluing the concept of local causality in classical physics.[4] I therefore argue from equations (2) and (3) that electric and magnetic fields may cause changes in each other, and that we should aim to teach this to students.

To illustrate the role of local causality in electromagnetism I will explain how the magnetic field due to an idealized uniform current density is established. A simple view is that the current causes the magnetic field. However, as I will show, this is an incomplete description as it neglects how the magnetic field evolves from zero.

We assume the system to be translationally symmetric in the *z* direction and cylindrically symmetric about the *z* axis, with radial coordinate *r* and azimuthal angle $\phi$. Hence the only spatial variation is radial. To avoid the complications of familiar systems such as wires we assume that a current density along the *z* direction is turned on instantaneously at time *t* = 0. [5] The current density is then fixed at $J_z$ uniformly everywhere with *r* < *R*, and zero elsewhere. I will refer to *r* = *R* as the current boundary. Starting from zero fields everywhere, equation (3) says that inside the current boundary, turning on the current causes an electric field $E_z$ to grow. This in turn generates an azimuthal magnetic field $B_\phi$. All other field components remain zero as there is nothing to cause them to grow. In cylindrical coordinates (*r*, $\phi$, *z*) equations (2) and (3) then reduce to:

$$\frac{\partial B_\phi}{\partial t} = \frac{\partial E_z}{\partial r}, \qquad \text{Faraday's law ;} \qquad (4)$$

$$\frac{\partial E_z}{\partial t} = c^2 \left( \frac{\partial B_\phi}{\partial r} + \frac{B_\phi}{r} \right) - \varepsilon_0^{-1} J_z, \quad \text{Ampere-Maxwell law.} \qquad (5)$$

Figure 1 shows a time-series of snapshots from a numerical solution of these equations.[6] The first two graphs clockwise from top left show the magnetic field (solid line) moving inwards and outwards from the current boundary. The last two show the establishment of the steady state magnetic field proportional to *r* inside the current boundary, and the development of the 1/*r* field outside the current boundary (note the change of scale on the *r* axis). The electric field inside the current boundary is growing in the first two graphs and is decaying towards zero in the second two (dashed line).

The causal story told by Figure 1 is as follows. The Ampere-Maxwell law, equation (5), says that turning on the current causes an initially spatially uniform electric field to evolve in the current region. Faraday's law, equation (4), says that the magnetic field stays zero where the electric field is spatially uniform. Initially, just outside the current boundary the electric field remains zero. Hence at the current boundary the curl of the electric field is non-zero, which causes the magnetic field to grow according to Faraday's law. The resulting non-uniform magnetic field, along with the current density, causes further evolution of the electric field. The net result is an electromagnetic field propagating inwards from the boundary causing the magnetic field to grow as it goes. Eventually the steady state magnetic field is established and the initial electric field is eliminated. The local cause of the final magnetic field in the current region is therefore to be found in local electric and magnetic fields as well as in the current. Outside the current region an electromagnetic pulse propagates outwards, eventually leaving behind a magnetic field varying inversely with the radial coordinate.

This analysis shows that the claim that only the current in the wire causes the magnetic field ignores its generation following Faraday's law and the Ampere-Maxwell law. At best it might be said that the current causes the magnetic field if timescales shorter than that light takes to cross the current region are ignored. On shorter time scales the local causes of the magnetic field are the spatial and temporal variations of the fields, together with the current. The initial growth of the magnetic field is due to an electromagnetic wave propagating in from the current boundary, where it was generated by the spatial discontinuity in the current density.

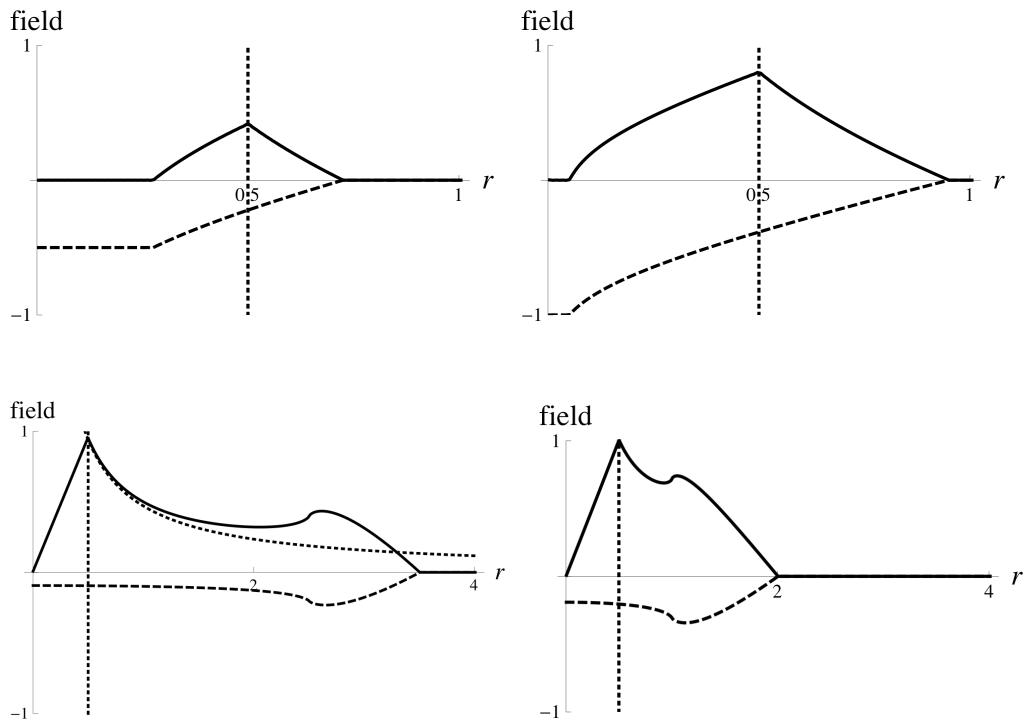

Figure 1. Graphs of the magnetic field $B_\phi$ (solid line) and electric field $E_z$ (dashed line) as a function of radial distance $r$ from the center of the circular current region, obtained from a numerical solution of equations (4) and (5). In the last two graphs the radial scale is expanded by a factor of four. The current boundary is at $R = 0.5$ m, indicated by the vertical dotted line. The radial distance is in meters and the magnetic and electric fields are normalized to their maxima of 0.033 T and 17 MV/m. The graphs are a time sequence; clockwise from top left they are for: $t = 0.75$ ns ($ct = 0.225$ m), 1.5 ns ($ct = 0.45$ m), 5 ns ($ct = 1.5$ m), and 10 ns ($ct = 3$ m) after the switch on of the current at $t = 0$.[5] The current density is $J_z = 0.1$ MA/m². The dotted curve in the last graph shows the expected asymptotic inverse radial variation of $B_\phi$.

What then should we teach students? I have argued that we should teach that spatially varying magnetic (electric) fields cause electric (magnetic) fields to change in time. As Hill argues, we should not teach that temporally changing magnetic fields cause electric fields.[1] He advocates teaching the retarded causes from charges and currents.[1,2] This has the advantage that students are comfortable with charges and currents as the origin of static fields. However such an explanation leaves unexamined how the cause propagates across space to produce its final effect. Retarded causes are not local causes.

The integral form of Maxwell's equations are usually taught first. However a locally causal explanation of classical electrodynamics requires their differential form. The integral form are not locally causal because they describe the behavior of fields over extended regions. As explained by Hill,[1,2] they describe field correlations, and I agree that this is what we should teach students until they are ready for the locally causal theory.